\documentclass[aps,nofootinbib,prd,eqsecnum,twocolumn,showpacs,showkeys,preprintnumbers]{revtex4-1}
\usepackage{graphicx}
\usepackage{amsmath}
\usepackage{amsfonts}
\usepackage{amssymb}
\usepackage{color}
\usepackage{bm}
\usepackage{float}
\usepackage{mathrsfs}
\usepackage{epstopdf}
\usepackage{url}
\usepackage{footnote}
\usepackage{textcomp}
\usepackage{placeins}
\usepackage{soul}

\usepackage[normalem]{ulem}
\usepackage{esint}
\usepackage[unicode=true, pdfusetitle,
 bookmarks=true,bookmarksnumbered=false,bookmarksopen=false,
 breaklinks=false,pdfborder={0 0 1},backref=false,colorlinks=false]{hyperref}
\usepackage{multirow}
\usepackage{pifont}
\usepackage{times}
\usepackage[english]{babel}

\usepackage{float}
\usepackage{enumerate}
\usepackage{hyperref}
\usepackage{tabularx}
\makeatletter
\newcommand{\stkout}[1]{\ifmmode\text{\sout{\ensuremath{#1}}}\else\sout{#1}\fi}

\newcolumntype{L}[1]{>{\hsize=#1\hsize\raggedright\arraybackslash}X}%
\newcolumntype{R}[1]{>{\hsize=#1\hsize\raggedleft\arraybackslash}X}%
\newcolumntype{C}[1]{>{\hsize=#1\hsize\centering\arraybackslash}X}%
\newcommand*\patchAmsMathEnvironmentForLineno[1]{%
 \expandafter\let\csname old#1\expandafter\endcsname\csname #1\endcsname
 \expandafter\let\csname oldend#1\expandafter\endcsname\csname end#1\endcsname
 \renewenvironment{#1}%
   {\linenomath\csname old#1\endcsname}%
   {\csname oldend#1\endcsname\endlinenomath}}% 
\newcommand*\patchBothAmsMathEnvironmentsForLineno[1]{%
 \patchAmsMathEnvironmentForLineno{#1}%
 \patchAmsMathEnvironmentForLineno{#1*}}%
\AtBeginDocument{%
\patchBothAmsMathEnvironmentsForLineno{align}%
\patchBothAmsMathEnvironmentsForLineno{flalign}%
\patchBothAmsMathEnvironmentsForLineno{alignat}%
\patchBothAmsMathEnvironmentsForLineno{gather}%
\patchBothAmsMathEnvironmentsForLineno{multline}%
}
\begin{document}

\title{Smoothing the $H_0$ tension with a dynamical dark energy model}
\author{Safae Dahmani}
\email{dahmani.safae.1026@gmail.com}
\author{Amine Bouali}
\email{a1.bouali@ump.ac.ma}
\author{Imad El Bojaddaini}
\email{i.elbojaddaini@ump.ac.ma}
\author{Ahmed Errahmani}
\email{ahmederrahmani1@yahoo.fr}
\author{Taoufik Ouali}
\email{t.ouali@ump.ac.ma}

\date{\today }
\affiliation{
Laboratory of Physics of Matter and Radiation, Mohammed I University, BP 717, Oujda, Morocco}

\begin{abstract}
The discrepancy between Planck data and direct measurements of the current expansion rate
$H_0$ and the matter fluctuation amplitude $S_8$ has become one of the most intriguing puzzles in cosmology nowadays.  The $H_0$ tension has reached $4.2\sigma$ in the context of standard cosmology i.e $\Lambda$CDM. Therefore, explanations to this issue are mandatory to unveil its secrets. Despite its success, $\Lambda$CDM is unable to give a satisfying explanation to the tension problem. Unless some systematic errors might be hidden in the observable measurements, physics beyond the standard model of cosmology must be advocated. In this perspective, we study a phantom dynamical dark energy model as an alternative to $\Lambda$CDM in order to explain the aforementioned issues. This phantom model is characterised by one extra parameter, $\Omega_{pdde}$, compared to $\Lambda$CDM. We obtain a strong positive correlation between $H_0$ and $\Omega_{pdde}$, for all data combinations.  Using Planck18 measurements together with BAO and Pantheon, we find that the $H_0$ and  the $S_8$ tensions are $3\sigma$ and $2.6\sigma$, respectively. By introducing a prior on the absolute magnitude, $M_B$, of the SN Ia, the $H_0$ tension decreases to $2.27\sigma$ with $H_0 = 69.76_{-0.82}^{+0.75}$ km s$^{-1}$ Mpc$^{-1}$ and the $S_8$ tension reaches the value $2.37\sigma$ with $S_8 =0.8269_{-0.012}^{+0.011}$.
\end{abstract} 
%%%%%%%%%%%%%%%%%%%%%%%%%%%%%%%%%%%%%%%%%%%%%%%%%%%%%%%%%%%
\maketitle
\section{introduction}

Supernova Type Ia (SN Ia) observation  \cite{SN1, SN2} reports an unexpected  cosmic acceleration of the expansion of the current Universe. This observation was corroboratted latter by other observations such as the
cosmic microwave background (CMB) \cite{CMB1, CMB2}, the large scale structure  \cite{LS1, LS2} and the baryonic acoustic oscillations (BAO) \cite{BAO1, BAO2}.  The standard model of cosmology successfully describes this late time cosmic acceleration by introducing a new exotic component in the budget of the Universe dubbed dark energy (DE). In the context of the standard model of cosmology, called $\Lambda$CDM, the major part of the content of the Universe is dominated by DE which is in the form of a cosmological constant, $\Lambda$,  and the cold dark matter (CDM).  In addition, various observational data give preference to $\Lambda$CDM for a vast range of redshifts $z$ \cite{L1, L2, PL, SU1, SU2}. However, this model faces many problems, among them the ``Hubble tension'', related to the current Hubble rate $H_0$ and the $\sigma_8$ tension due to the matter fluctuation amplitude.\\
%%%%%%%%%%%%%%%%%%%%%%%%%%%%%%%%%%%%%%%%%%%%%%%%%%%%%%%%%%%%%%%%
The Hubble tension appears when comparing the value measured indirectly by calibrating theoretical models in the early Universe i.e. at high-redshift and that measured directly using cosmological distances and redshifts by observing space objects. Generally, the value obtained at high-redshifts is lower than that obtained at low-redshifts. The value predicted at high-redshifts i.e. by cosmic microwave background measurements assuming the $\Lambda$CDM model, is $H_0=67.37\pm 0.54$ km s$^{-1}$ Mpc$^{-1}$ \cite{PL} while the one determined by the Cepheid calibrated supernovae Ia, is $H_0=73.2\pm 1.3$ km s$^{-1}$ Mpc$^{-1}$ \cite{R21}. It is clear that there is significant discrepancy between these values qualified as a tension. This tension is at about $4.2\sigma$ level. Recent studies  have shown that this tension depends directly on the SN Ia absolute magnitude, $M_B$ \cite{EF, MB, MB1, MB2, MB3}. In fact the SH0ES project measures the absolute peak magnitude ($M_B=-19.244\pm 0.037$ mag \cite{MB}) of SN Ia, while the value of $H_0$ can be estimated by the magnitude-redshift relation of SN Ia in the range $z\in[0.023, 0.15]$ \cite{Scolnic_2018}. The same studies indicated that to test any model that modifies the late-time of the Universe, it is necessary to use a prior on the absolute magnitude of supernovae type Ia, $M_B$, instead of using the prior on $H_0$ from SH0ES for a correct statistical analysis and to avoid misleading results. \\
On the other hand, the tension between the value of the matter ﬂuctuation amplitude  $\sigma_8$ found by CMB measurements and that from large-scale observations in the late Universe rises another problem in $\Lambda$CDM. The parameter that quantifies the matter fluctuations is defined by $S_8 = \sigma_8 \sqrt{(\Omega_{m,0}/0.3)},$ representing a combination of $\sigma_8$ and the matter density $\Omega_{m,0}$ at the present time. Constraints from Planck18 and those from local measurements are in tension at more than $2\sigma$. Indeed, while the constrained $S_8$ from Planck18 data is $S_8 = 0.832 \pm 0.013$ \cite{PL}, smaller values are found from local measurements, e.g. $S_8=0.762^{+0.025}_{-0.024}$ obtained by KV450 (KiDS+VIKING-450) and DES-Y1 (Dark Energy Survey Year 1) combined \cite{DESY1}.\\
This discrepancy could be an evidence of new physics beyond the standard model of cosmology \cite{FR, SM, SM1}. Several theoretical approaches have been proposed to solve these tension problems, such as extensions of the $\Lambda$CDM model, DE–DM interactions and decaying DM \cite{LCDM1,I1, I2, I3, I4, I5, I6, DM1, 19, 20, s8, t8, S88, T9, T10, T11, T12}. These approaches have also shown that changing the properties of DE e.g. by introducing the early Dark Energy \cite{early, early2, early3} and the phantom Dark Energy where the equation of state (EoS) parameter is slightly less than $-1$, can increase the value of $H_0$ and consequently can alleviate the Hubble tension compared to $\Lambda$CDM \cite{P1, P2, P3, P4,P10, P6, P8, P9, P11, phan1, phan2, P111}.\\

These conclusions motivated us to address these issues in the context of a particular dynamical dark energy (DDE) model where the EoS and the energy density are given respectively by \cite{LSBR1}
\begin{equation}
p_{de}=-(\rho_{de}+\frac{\alpha}{3}),
\label{d1}
\end{equation}
and
\begin{equation}
\rho_{de}(z)=\rho_{de,0}-\alpha\ln{(1+z)},
\label{d2}
\end{equation}
where $\rho_{de,0}$ is the present DE density, $\alpha$ is a positive constant that distinguishes this model from $\Lambda$CDM. Hence $\rho_{de}$ tends to the standard cosmological constant $\Lambda$ at the present ($z=0$). This phantom dynamical dark energy  model induces an abrupt event in the future  where the dark energy density  dominates all  other forms of energy density. However, in the past, this dark energy density decreases and the energy density of dark matter dominates the budget of the Universe.\\
According to Eqs. (\ref{d1}) and (\ref{d2}) the EoS parameter is given by
\begin{equation}
\omega_{de}=-(1+\frac{\alpha}{3(\rho_{de,0}-\alpha\ln{(1+z)})}).
\label{d3}
\end{equation}
For positive values of $\alpha$, we get $\omega_{de}<-1$  and the model describes a phantom dark energy. For negative values of $\alpha$, $\omega_{de}>-1$, the model describes a quintessence dark energy and mimics $\Lambda$CDM  in the limit $\alpha\rightarrow 0$. In the following, we will focus on the phantom case i.e. $\alpha>0$.\\
The Friedmann equation of a Universe filled by CDM and DE can be written as \cite{LSBR1, LSBR2, LSBR3}
\begin{equation}
E^2(z)=\Omega_{m,0}(1+z)^3+\Omega_{de,0}-\Omega_{pdde}\ln{(1+z)},
\label{d4}
\end{equation}
where $E=\frac{H(z)}{H_0}$, $H_0$ is the current Hubble rate, $\Omega_{m,0}$ is the actual matter density, $\Omega_{de,0}=\frac{8\pi G}{3H^2_0}\rho_{de,0}$ and $\Omega_{pdde}=\frac{8\pi G}{3H^2_0}\alpha$ is a dimensionless parameter characterizing our phantom DDE model.\\
From Eq (\ref{d4}), the model predicts more dark matter  in the past as $z\rightarrow \infty$. However, in the future, this model is dominant by the DE and it is characterized by a particular behaviour. Indeed, its  Hubble rate $H$ diverges while its  derivative $\dot{H}$ remains constant. This abrupt event has been well studied in \cite{LSBR1, LSBR2, LSBR3, LSBR4, LSBR5, LSBR6, LSBR7, LSBR8} and dubbed as  Little Sibling of the Big Rip  since it smooths the big rip singularity in the future.\\
In this paper, we study the effect of this  phantom dynamical dark energy model (PDDE) on  both tensions, namely the  $H_0$ and $S_8$ tensions,  and  we compare the results with those of $\Lambda$CDM. To this aim, we perform a Markov Chain Monte Carlo (MCMC) \cite{MCMC} analysis, using different datasets.\\
This paper is organized as follows: in Sec. \ref{sec1}, we describe the methodology followed and the data used in our analysis. In Sec. \ref{sec2}, we present the results and  discussions,  while in Sec. \ref{sec4} we analyze the effect on the power spectrum. Finally, Sec. \ref{sec3} is dedicated to conclusions.
\section{Methodology and datasets}\label{sec1}
To constrain our theoretical model we employ the $\chi^2$ statistics\\
\begin{equation}
\chi^2=\frac{[\mathcal{P}_{obs}-\mathcal{P}_{th}]^2}{\sigma^2_{\mathcal{P}}},
\end{equation}
 where $\mathcal{P}_{obs}$, $\mathcal{P}_{th}$ and  $\sigma^2_{\mathcal{P}}$ indicate the observed values, the predicted values and the standard deviation, respectively. The model with  a small value of $\chi^2$ fits better the observational data and  is considered as the best. We also use the Akaike Information Criterion (AIC) \cite{AIC}, which is widely used in cosmology \cite{AIC1, AIC2} to compare cosmological models with different free parameters numbers 
\begin{equation}
AIC=-2\ln{(\mathcal{L}_{max})}+2N,
\end{equation}
where $\mathcal{L}$  is the likelihood and $N$ is the number of free parameters. The model with a small value of AIC is the most supported by observational data. In this work, we calculate $\bigtriangleup AIC=AIC_{PDDE}-AIC_{\Lambda CDM}$ and we consider $\Lambda$CDM as the reference model i.e. $\bigtriangleup AIC=0$. Furthermore, a positive (negative) value of  $\bigtriangleup AIC$ indicates that $\Lambda CDM$ (PDDE) is the most preferred model by observational data. \\

To run the MCMC \cite{MCMC} we use the MontePython code \cite{MP}, which interfaces with CLASS \cite{CLASS} in which we have implemented our PDDE fluid. We consider 7-dimensional parameters space, consisting of six standard cosmological parameters $\omega_b$, $\omega_{cdm}$, $H_0$, $n_s$, $\tau_{reio}$ and $\ln{(10^{10}A_{s })}$ which correspond to the physical densities of baryons and CDM, the Hubble constant, the scalar spectral index, the optical depth and the power spectrum amplitude, respectively plus the additional parameter $\Omega_{pdde}$ characterizing our PDDE model. The priors of these free parameters are mentioned in Table \ref{1}. To avoid non-adiabatic instabilities at the perturbation evolution, we employ the Parameterized Post-Friedmann (PPF) \citep{Fang_2008} approach.\\

In this work, we use the following observational data:\\
\textbf{Planck18}: The CMB temperature measurements  (\textit{low-$\ell$ TT}) and polarization (\textit{low-$\ell$ EE}) at low multipoles $2\leqslant\ell\leqslant 29$.  We also use temperature and polarization combined (\textit{high-$\ell$ TT TE EE}) at higher multipoles $30\leqslant\ell\leqslant 2500$. In addition we use the lensing constraint \cite{PL}.\\
\textbf{BAO}: The Baryon Acoustic Oscillation measurements at different redshifts $z$, BOSS\_DR12 from the CMASS (at $z=0.57$) and LOWZ galaxies (at $z=0.32$) ~\cite{Alam_2017}, 6dFGS (at $z=0.106$)~\cite{Beutler_2011} and SDSS DR7 (at $z=0.15$)~\cite{Ross_2015}.\\
\textbf{Pantheon}:  The luminosity distance from 1048 Supernovae Type Ia (SN Ia) in the range $z\in[0.01, 2.3]$~\cite{Scolnic_2018}. The SN Ia data directly give measures of $m_b(z)$ for each $z$, where $m_b(z)$ is the apparent magnitude. For a given cosmological model this parameter can be calculated theoretically by
\begin{equation}
m_b(z)=5log_{10}[\frac{d_L(z)}{Mpc}]+M_B+25,
\end{equation}

where,  $d_L(z)=(1+z)\int_{0}^{z}\frac{cdz'}{H(z')}$ is the luminosity distance and  $M_B$ is the absolute magnitude which will be considered as a free parameter in our analysis.\\
\textbf{Prior on $\bm{M_B$}}: The SN measurements from the SH0ES project  give a Gaussian prior on the absolute magnitude as $M_B = -19.244\pm 0.037$ mag \cite{MB}.\\

The total $\chi^2$ of the combined data is
\begin{equation}
\chi^2_{tot}=\chi^2_{Planck18}+\chi^2_{BAO}+\chi^2_{Pantheon}+\chi^2_{M_B}.
\end{equation}
\begin{table}[!htp]
\centering
{\caption{A prior imposed on different parameters for the $\Lambda$CDM and  PDDE models}\label{1}}
\begin{tabular}{c|c}
\hline
\multicolumn{1}{c|}{\bf Parameters} & \multicolumn{1}{c}{\bf Prior}\\
\hline      
$\Omega_{\textrm{b}} h^{2}$   & [$0.005, 0.1$]
 \\[0.1cm]
$\Omega_{\textrm{c}} h^{2}$ &[$0.01, 0.99$]
 \\[0.1cm]
 
$\Omega_{\textrm{pdde}}$ &[$0,$ $1$]
 \\[0.1cm]
$H_0$ &[$40,100$]
 \\[0.1cm]
$\tau_{reio}$ &[$0.001,0.8$]
 \\[0.1cm]
$n_{s }$ &[$0.8,1.2$]
 \\[0.1cm]
$\ln{(10^{10}A_{s })}$ &[$2,4$]
\\[0.1cm]
$M_B$ & default prior\footnote{For the absolute magnitude parameter we used MontePython v3.5 default prior.}
 \\[0.1cm]
\hline
\end{tabular}
\end{table}
\section{Results and discussions}\label{sec2}
We perform an MCMC analysis to obtain constraints on cosmological parameters of the PDDE model and compare them with those of $\Lambda$CDM. First of all, we employ the PDDE model with three different data combinations, namely Planck18, Planck18+BAO and Planck18+BAO+Pantheon, in order to make a comparison with $\Lambda$CDM and get a general insight of the analysis. In the second analysis, we include a prior on $M_B$ from SH0ES to Planck18+BAO+Pantheon datasets.
\subsection{Planck18, BAO and Pantheon datasets.}
\begin{table*}[!htp]
\centering
{\caption{Summary of the mean$\pm1\sigma$ of the cosmological parameters for the  $\Lambda$CDM and  PDDE models, using Planck18, Planck18+BAO and Planck18+BAO+Pantheon datasets.}\label{2}}
\begin{tabular}{c|cc|cc|cc}
\hline
\hline
\multicolumn{1}{c|}{Data} & \multicolumn{2}{c|}{Planck18\footnote{We used the ``lite'' version of high-$\ell$ likelihood.}} & \multicolumn{2}{c|}{Planck18+BAO}&\multicolumn{2}{c}{Planck18+BAO+Pantheon}\\
\hline
\multicolumn{1}{c|}{Model} & \multicolumn{1}{c}{$\Lambda$CDM} & \multicolumn{1}{c|}{PDDE}& \multicolumn{1}{c}{$\Lambda$CDM} & \multicolumn{1}{c|}{PDDE} & \multicolumn{1}{c}{$\Lambda$CDM} & \multicolumn{1}{c}{PDDE}\\
\hline
$100\Omega_{\textrm{b}} h^{2}$   &$2.238_{-0.017}^{+0.016}$   & $2.236_{-0.016}^{+0.015}$ & $2.24_{-0.013}^{+0.014}$&$2.236\pm{0.015}$& $2.241\pm{0.014}$&$ 2.236\pm{0.015}$
 \\[0.1cm]
$\Omega_{\textrm{c}} h^{2}$  &$0.1199\pm0.0013$  & $0.1203_{-0.0013}^{+0.0012}$  &  $0.1196_{-0.00098}^{+0.00099}$ &$0.12\pm{0.0011}$& $0.1196_{-0.001}^{+0.00096}$& $0.12\pm{0.001}$
 \\[0.1cm]
$\Omega_{\textrm{pdde}}$  $(95\%)$ &-  &  $<0.06401$& -& $<0.2024$&-&$<0.1376$
  \\[0.1cm]
$H_0$ [km s$^{-1}$ Mpc$^{-1}$] & $67.93_{-0.63}^{+0.58}$   & $67.66_{-0.67}^{+0.65}$ & $68.07\pm{0.45}$&$ 69.13_{-1.1}^{+0.79} $& $68.09_{-0.45}^{+0.46}$& $68.76_{-0.76}^{+0.59}$
\\[0.1cm]
$\tau_{reio}$  &$0.054\pm{0.0081}$ & $0.05496_{-0.0079}^{+0.0072}$&  $0.0545_{-0.0077}^{+0.0076}$&$ 0.0528_{-0.008}^{+0.0075}$&$0.055_{-0.0079}^{+0.0074}$& $0.053\pm{0.0076}$
\\[0.1cm]
$n_{s }$   &$0.965\pm{0.0046}$&   $0.9647_{-0.0042}^{+0.0043}$  &  $0.9661_{-0.0041}^{+0.0039}$& $0.964\pm{0.0041}$& $0.9663_{-0.0038}^{+0.0039}$ &$0.965_{-0.0039}^{+0.0042}$ 
\\[0.1cm]
$\ln{(10^{10}A_{s })}$  &$3.044_{-0.016}^{+0.015}$ &  $3.046_{-0.016}^{+0.015}$  &  $3.044_{-0.016}^{+0.015}$&$3.041\pm{0.015}$& $3.045\pm{0.015}$& $3.042\pm{0.015}$
\\[0.1cm]
$\Omega_{\textrm{m}}$   & $0.308_{-0.008}^{+0.0081}$& $0.3131_{-0.0091}^{+0.0079}$ &  $0.3065_{-0.0061}^{+0.0058}$&$0.2982_{-0.0082}^{+0.0094}$&$0.3063_{-0.0062}^{+0.0057}$ & $0.3013_{-0.0066}^{+0.007}$
 \\[0.1cm]
$\sigma_8$  &$0.823\pm{0.0066}$ & $0.8155_{-0.0074}^{+0.0065}$&  $0.8226\pm{0.0066}$& $0.835_{-0.013}^{+0.0094}$ &$0.8227_{-0.0066}^{+0.0062}$ & $0.831_{-0.01}^{+0.0082}$ 
\\[0.1cm] 
$S_8$  &$0.834\pm{0.014}$ & $0.833\pm{0.014}$&  $0.831\pm{0.011}$& $0.832_{-0.012}^{+0.011}$ & $0.8312_{-0.012}^{+0.011}$ &$0.832\pm{0.011}$
\\[0.1cm]
 $M_B$ [mag]  &- & -&  -& - &  $-19.41_{-0.012}^{+0.013}$  & $-19.39_{-0.017}^{+0.015}$
\\[0.1cm]
\hline
$H_0$ tension\footnote{To calculate the tension between two values of $H_0$ obtained from different data (d1, d2), we use the following expression \cite{H1, H2}: $T(H_0)=\vert H_{0(d1)}-H_{0(d2)}\vert/(\sqrt{\sigma(H_{0(d1)})^2+\sigma(H_{0(d2)})^2})$, where $H_0$ and $\sigma$ are the mean and the variance of the posterior of Hubble rate (the same for $M_B$ and $S_8$).}  &$3.6\sigma$  &$3.7\sigma$&  $3.72\sigma$&$2.531\sigma$&$3.7\sigma$&$3\sigma$
\\[0.1cm] 
$M_B$ tension  & - & - & -&-&$4.2\sigma$ & $3.6\sigma$
\\[0.1cm]  
$S_8$ tension  &$2.6\sigma$  & $2.4\sigma$ & $2.6\sigma$&$2.6\sigma$&$2.5\sigma$&$2.6\sigma$
\\[0.1cm] 
\hline
\hline
\end{tabular}
\end{table*}
\begin{figure*}[!htp]
\centering
\includegraphics[width=17cm,height=15cm]{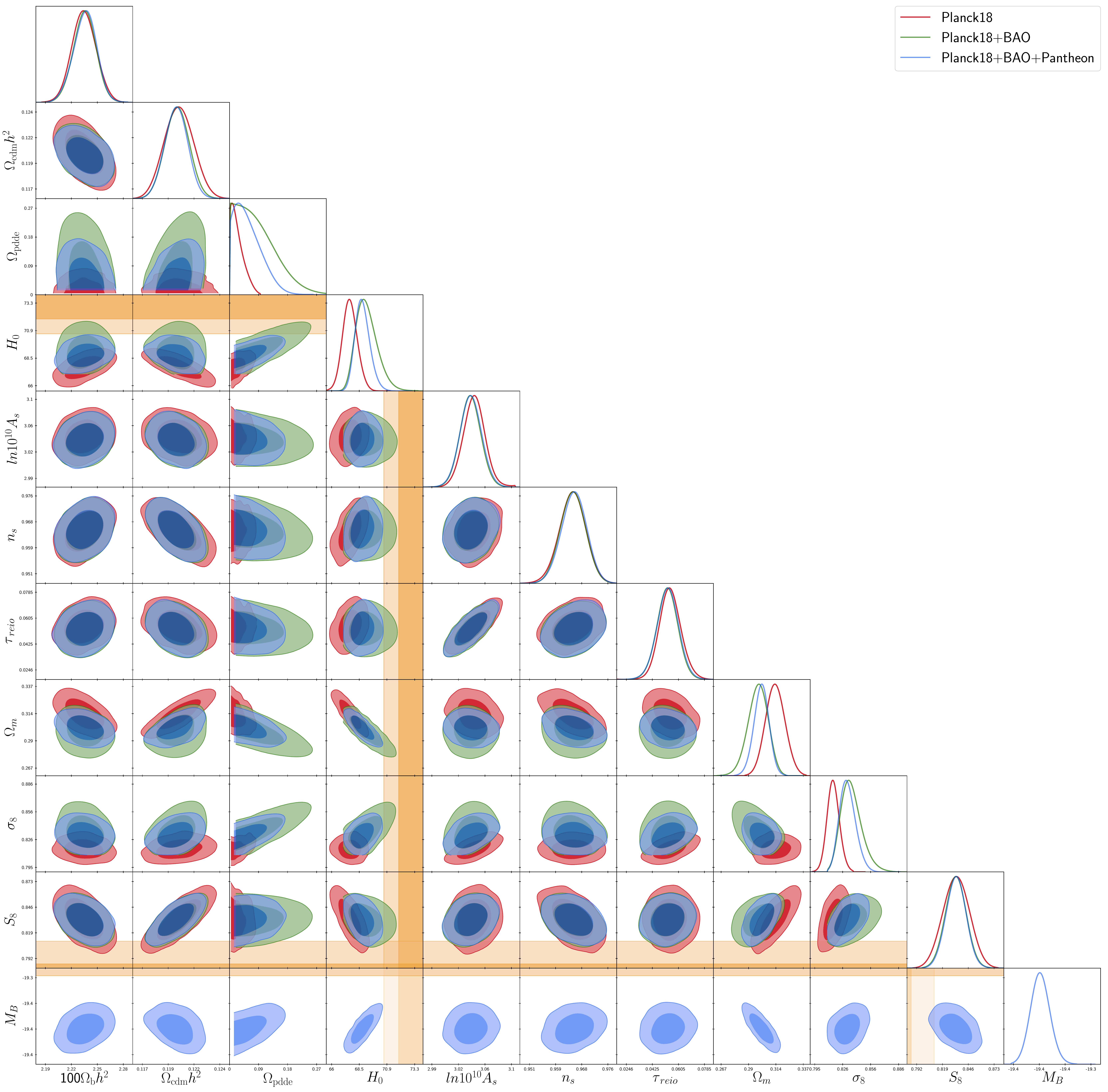} 
\caption{The 2D and 1D posterior distributions at 68.3\% and 95.4\% C.L. for the PDDE model using different combinations of data (Planck18, Planck18+BAO and Planck18+BAO+Pantheon). The local measurement of $H_0=73.2\pm1.3 $km s$^{-1}$ Mpc$^{-1}$ and $S_8=0.762^{+0.025}_{-0.024}$ obtained by R20 and KV450+DES-Y1 respectively, are represented by the orange band.}
\label{f1}
\end{figure*}
\begin{figure*}
\centering
\includegraphics[width=8.5cm,height=7.5cm]{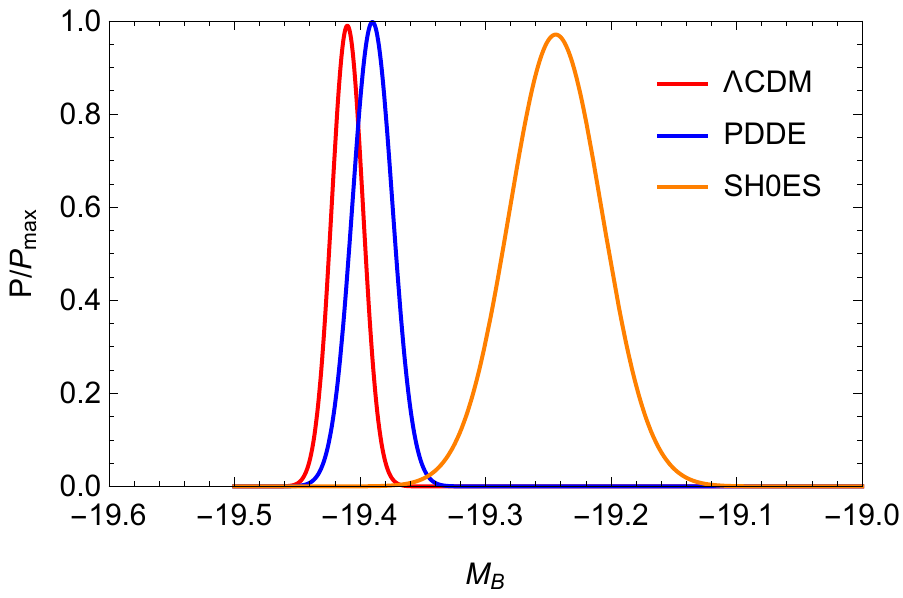}\hfill
\includegraphics[width=8.5cm,height=7.5cm]{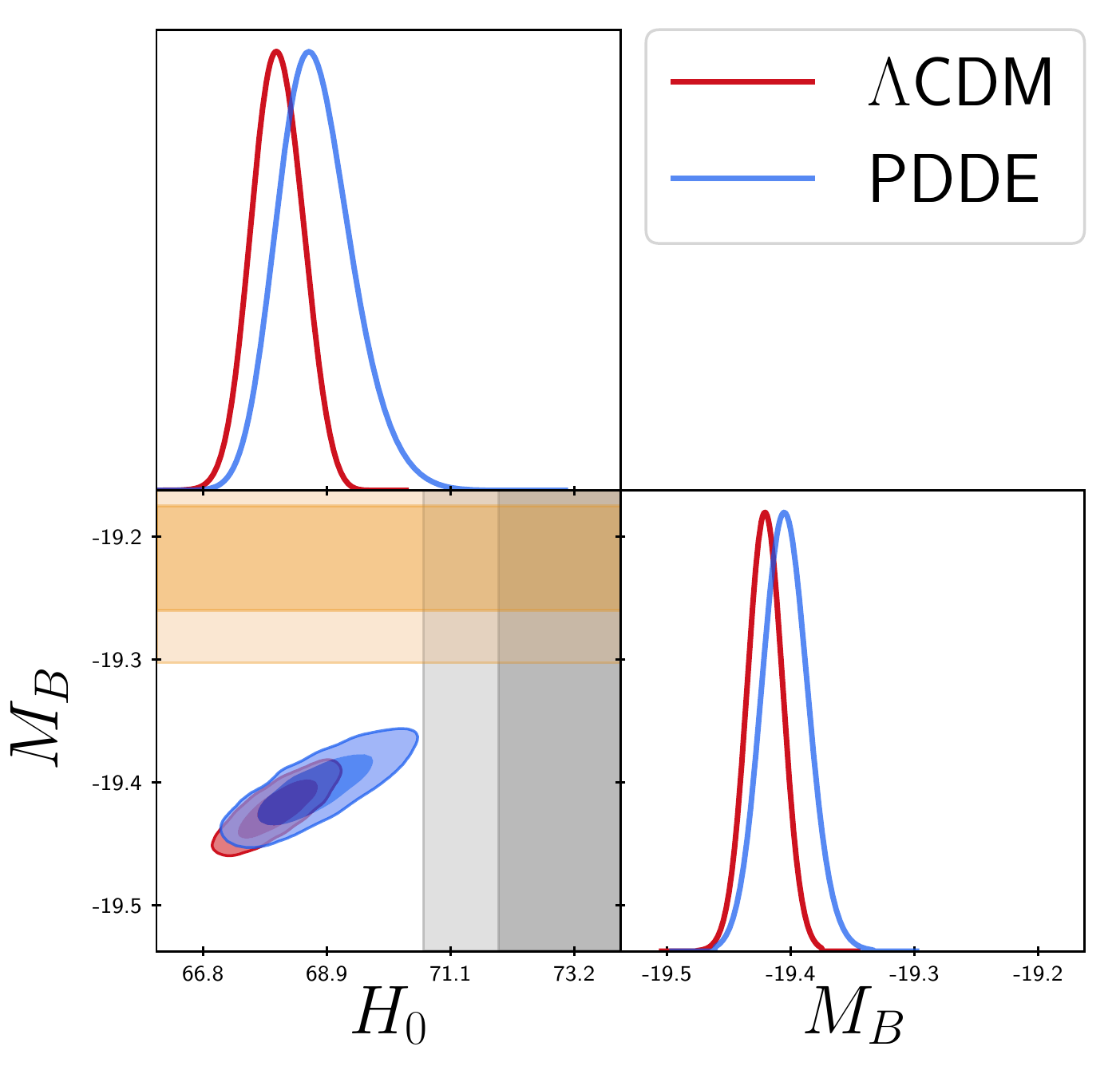}
\caption{The left panel shows 1D posterior distributions for the absolute magnitude, $M_B$. The right panel shows 68\% and 95\% constraints on ($H_0$, $M_B$) plan using Planck18+BAO+Pantheon datasets. The local measurement of $H_0=73.2\pm1.3 $km s$^{-1}$ Mpc$^{-1}$ and $M_B = -19.244\pm 0.037$ mag  obtained by SH0ES, are represented by the grey band and the orange band respectively. }
\label{fmb}
\end{figure*}

\begin{table*}[!htp]
\centering
{\caption{The best-fit $\chi^2$ per experiment for the $\Lambda$CDM and  PDDE models.}\label{3}}
\begin{tabular}{c|cc|cc|cc}
\hline
\multicolumn{1}{c|}{Datasets}    & \multicolumn{2}{c|}{Planck18}& \multicolumn{2}{c|}{Planck18+BAO} & \multicolumn{2}{c}{Planck18+BAO+Pantheon} \\
\hline  

\multicolumn{1}{c|}{Model}  & \multicolumn{1}{c}{ $\Lambda$CDM} & \multicolumn{1}{c|}{ PDDE}& \multicolumn{1}{c}{ $\Lambda$CDM} & \multicolumn{1}{c|}{ PDDE}& \multicolumn{1}{c}{ $\Lambda$CDM} & \multicolumn{1}{c}{ PDDE} \\[0.1cm]

\hline 
Planck\_high-$\ell$\_TTTEEE\_lite      & $ 583.41$ &$584.88$  &$583.96$  & $584.26$ &$ 583.5$&$ 583.96$ 
\\[0.1cm]
Planck\_low-$\ell$\_EE   &$396.23$ &$395.68$   &$395.98$  & $395.86$ &$ 396.26$&$395.84$   \\[0.1cm]
Planck\_low-$\ell$\_TT   & $23.44$ &$23.45$   &$23.27$  & $ 23.22$ &$ 23.36$&$ 23.43$  \\[0.1cm]
Planck\_lensing   & $ 8.78$ &$ 8.71$  &$8.81$  & $8.8$ &$ 8.801$&$8.81$    \\[0.1cm]
bao\_boss\_dr12      & - &-   &$3.73$  & $3.69$ &$ 3.88$&$3.92$   \\[0.1cm]
bao\_smallz\_2014      & - &-  &$1.48$  & $1.53$ &$  1.41$&$ 1.43$  \\[0.1cm]
Pantheon      & - &-  &-  & - &$ 1025.84$&$1025.80$  \\[0.1cm]
\hline
\hline

$\chi^2_{tot}$    & $ 1011.88$ &$1012.81$   &$1017.25$  & $1017.39$ &$ 2043.09$&$2043.23$  
\\[0.1cm]
$\bigtriangleup \chi^2_{tot}$     & $0$ &$+0.93$   &$0$  & $+0.14$ &$0$&$+0.14$  
\\[0.1cm]
$AIC$       & $1029.88$ &$1032.81$    &$1035.25$  & $1037.39$ &$ 2063.09$&$2065.23$  
\\[0.1cm]
$\bigtriangleup AIC$      & $0$ &$+2.93$    &$0$  & $+2.14$ &$ 0$&$+2.14$ 
\\[0.1cm]

\hline        
\end{tabular}
\end{table*}
Table \ref{2} shows the mean values and their corresponding errors at $68\%$ C.L.  for  all considered parameters using  Planck18, Planck18+BAO and Planck18+BAO+Pantheon. Fig. \ref{f1} shows the 2D and 1D posterior distributions for the PDDE model for the aforementioned datasets.\\
Using Planck18 data alone,  we get  $H_0=67.93_{-0.63}^{+0.58}$ km s$^{-1}$ Mpc$^{-1}$ for $\Lambda$CDM and $H_0=67.66_{-0.67}^{+0.65}$, for the PDDE model. The latter is about $0.2\sigma$ away from the value obtained by $\Lambda$CDM. Therefore, we can deduce that the PDDE model has the same behavior as the standard model, $\Lambda$CDM, at $z>1000$. As a result, we obtain a $H_0$ tension which is higher than $3\sigma$ with the local measurement of R20 i.e. $H^{R20}_0=73.2\pm 1.3$ km s$^{-1}$ Mpc$^{-1}$ for both models i.e. $3.6\sigma$ ($3.7\sigma$) for $\Lambda$CDM (PDDE). We can also see a $2\sigma$ ($95\%$) upper bound on $\Omega_{pdde}$ i.e. $<0.06401$. However, when we add the BAO data the upper bound  of $\Omega_{pdde}$ increases to $<0.2024$. We also see a small increase of $H_0$ by $0.2\%$ for $\Lambda$CDM i.e. $H_0=68.07\pm{0.45}$ km s$^{-1}$ Mpc$^{-1}$ with a tension of $\sim 3.7\sigma$  and by $2.12\%$ i.e. $H_0=69.13_{-1.1}^{+0.79}$ km s$^{-1}$ Mpc$^{-1}$ for the PDDE model. Therefore, the tension of $H_0$ decreased to  $\sim 2.5\sigma$ for the PDDE model compared to $H_0^{R20}$.  The significant difference between $H_0^{\Lambda CDM}$ and $H_0^{PDDE}$ tensions is actually not enough to come out with conclusive results about the $H_0$ tension because analyzing this tension in light of any late-time model like PDDE should necessarily involve analyzing the Pantheon SN Ia sample \cite{EF}. Adding Pantheon data to Planck18+BAO datasets, we observe a decrease in the upper limit of  $\Omega_{pdde}$ and in the value of $H_0$ to $<0.1376$ and $68.76_{-0.76}^{+0.59}$ km s$^{-1}$ Mpc$^{-1}$, respectively for the PDDE model. The tension in this case is still significant with $3\sigma$ for PDDE. However, this value is less than that of $\Lambda$CDM that gives $3.7\sigma$.  These conclusions can be justified by the positive correlation observed in the ($\Omega_{pdde}$, $H_0$) plan  as can be seen in Fig. \ref{f1}. We also compare the absolute magnitude, $M_B$, value with SH0ES calibration i.e. $M_B = -19.244\pm 0.037$ mag, we notice that our model can considerably smooth the $M_B$ tension, where the tension with SH0ES calibration is at about $3.6\sigma$, while for $\Lambda$CDM  is at about $4.2\sigma$ (see Fig. \ref{fmb}).
 On the other hand, in the context of the PDDE model, we obtain $ \sigma_8 = 0.8155_{-0.0074}^{+0.0065}$  and $\Omega_m= 0.3131_{-0.0091}^{+0.0079}$, which leads to $S_8=0.833\pm{0.014}$ and the $S_8$ tension is at $2.4\sigma$ for PDDE model compared to KV450+DES-Y1 Surveys i.e. $S_8=0.762^{+0.025}_{-0.024}$, and at $2.6\sigma$ for $\Lambda$CDM. Using Planck18+BAO and Planck18+BAO+Pantheon, we get $S_8=0.831\pm{0.011}$ ($S_8=0.832_{-0.012}^{+0.011}$) and $0.8312_{-0.012}^{+0.011}$ ($0.832\pm{0.011}$) for $\Lambda$CDM (PDDE), respectively. $S_8$ tension is at $2.6\sigma$ ($2.6\sigma$) and $2.5\sigma$ ($2.6\sigma$) for $\Lambda$CDM (PDDE).\\
Table \ref{3} shows the  $\chi^2$ for each data combination. Furthermore,  $\bigtriangleup\chi_{tot}^2=\chi_{tot}^{2 (PDDE)}-\chi_{tot}^{2 (\Lambda CDM)}$, $\bigtriangleup AIC=AIC^{PDDE}-AIC^{\Lambda CDM}$  and AIC are also shown in Table \ref{3}. Using Planck18 together with BAO and Pantheon datasets we get a positive value of $\bigtriangleup\chi_{tot}^2$ and $\bigtriangleup AIC$. The inclusion of these data gives preference to $\Lambda$CDM.\\
In the following, we will focus on the data combination Planck18+BAO+Pantheon as it is the only suitable combination to study the tension in the framework of the late-time model, PDDE.
\subsection{Adding $M_B$ prior.}
\begin{figure*}[!htp]
\centering
\includegraphics[width=17cm,height=15cm]{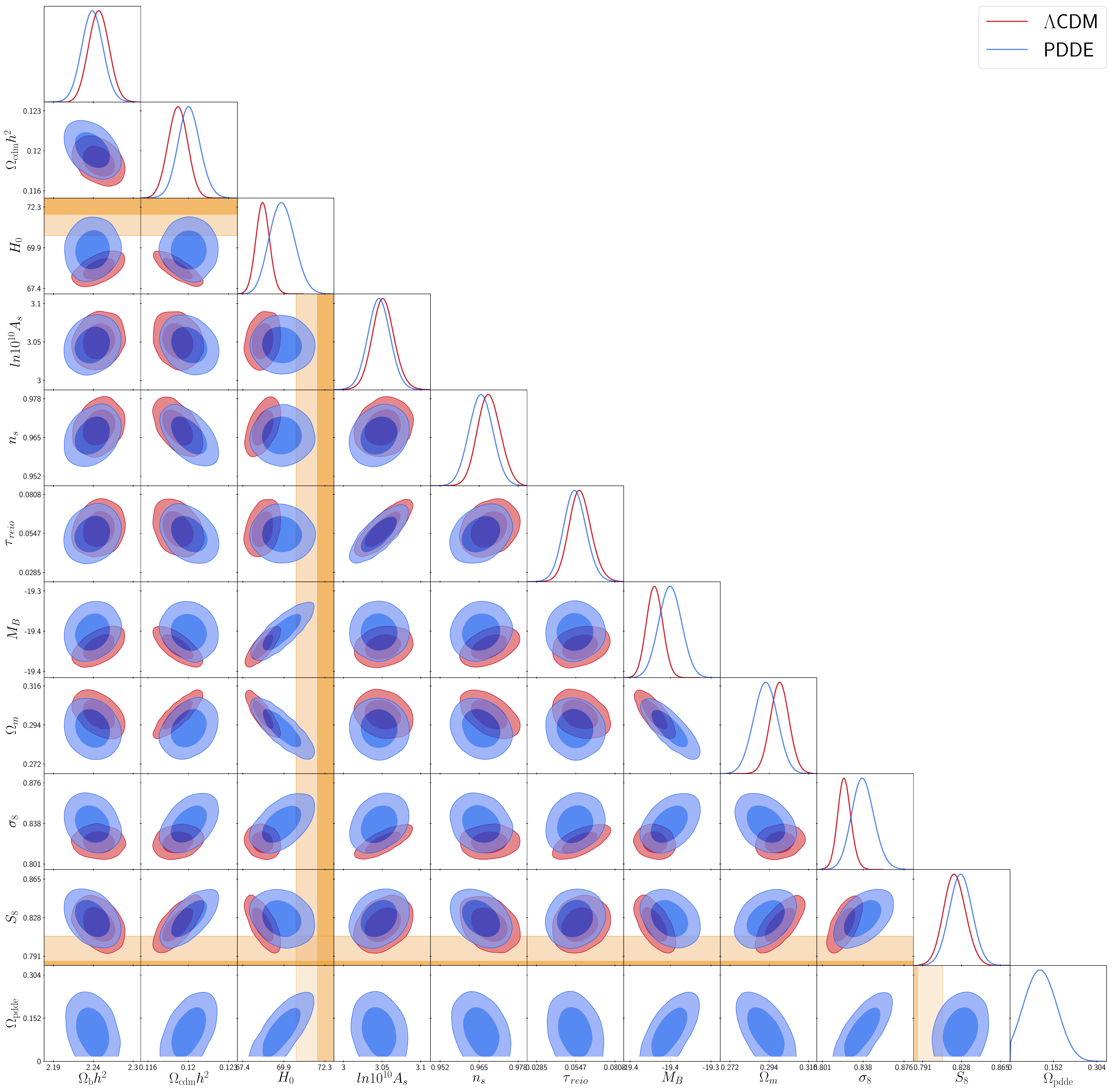} 
\caption{The 2D and 1D posterior distributions at 68.3\% and 95.4\% C.L. for the $\Lambda$CDM and PDDE models using Planck18+BAO+Pantheon+$M_B$ datasets. The local measurement of $H_0=73.2\pm1.3$ km s$^{-1}$ Mpc$^{-1}$ and $S_8=0.762^{+0.025}_{-0.024}$ obtained by SH0ES and KV450+DES-Y1 respectively, are shown by the orange band.}
\label{f3}
\end{figure*}
\begin{table*}[!htp]
\centering
{\caption{Summary of the mean$\pm1\sigma$ of cosmological parameters for the  $\Lambda$CDM and PDDE models,  using Planck18+BAO+Pantheon+$M_B$ datasets.}\label{5}}
\begin{tabular}{c|cc}
\hline
\hline
\multicolumn{1}{c|}{Data} &\multicolumn{2}{c}{Planck18+BAO+Pantheon+$M_B$}\\
\hline
\multicolumn{1}{c|}{Model} &\multicolumn{1}{c}{$\Lambda$CDM} & \multicolumn{1}{c}{PDDE}\\
\hline
$100\Omega_{\textrm{b}} h^{2}$   & $2.25\pm{0.014}$ &$2.242\pm{0.015}$
 \\[0.1cm]
$\Omega_{\textrm{c}} h^{2}$  &$0.1185_{-0.00098}^{+0.00093}$&$0.1196\pm{0.0011}$
 \\[0.1cm]
$\Omega_{\textrm{pdde}}$ &-&$0.1087_{-0.061}^{+0.052}$
  \\[0.1cm]
$H_0$ [km s$^{-1}$ Mpc$^{-1}$]&$68.58_{-0.44}^{+0.43}$ &$69.76_{-0.82}^{+0.75}$
\\[0.1cm]
$\tau_{reio}$ & $0.05758_{-0.0085}^{+0.0073}$& $0.05414_{-0.0081}^{+0.0077}$ 
\\[0.1cm]
$n_{s }$   & $0.9688_{-0.004}^{+0.0039}$&  $0.9661_{-0.0042}^{+0.0041}$ 
\\[0.1cm]
$\ln{(10^{10}A_{s })}$  &$3.049_{-0.016}^{+0.015}$& $3.043_{-0.016}^{+0.015}$
\\[0.1cm]

$\Omega_{\textrm{m}}$   & $0.3_{-0.0056}^{+0.0054}$&  $0.292\pm{0.0071}$
 \\[0.1cm]
$\sigma_8$ &$0.8213_{-0.0068}^{+0.0064}$ &$0.8382_{-0.012}^{+0.011}$ 
\\[0.1cm] 
$S_8$  &$0.8212\pm{0.011}$& $0.8269_{-0.012}^{+0.011}$
\\[0.1cm]
$M_B$ [mag]  & $-19.39\pm{0.012}$&  $-19.37_{-0.018}^{+0.017}$ 
\\[0.1cm]
\hline
$H_0$ tension &$3.36\sigma$& $2.27\sigma$
\\[0.1cm] 
$M_B$ tension  & $3.7\sigma$ & $3.06\sigma$ 
\\[0.1cm]
$S_8$ tension &$2.16\sigma$& $2.37\sigma$
\\[0.1cm] 
\hline
\hline
\end{tabular}
\end{table*}
\begin{table*}[!htp]
\centering
{\caption{The  $\chi^2$ per experiment for the  $\Lambda$CDM and PDDE models}\label{6}}
\begin{tabular}{c|cc}
\hline
\multicolumn{1}{c|}{Dataset}   & \multicolumn{2}{c}{Planck18+BAO+Pantheon+$M_B$} \\
\hline  

\multicolumn{1}{c|}{Model}  & \multicolumn{1}{c}{ $\Lambda$CDM} & \multicolumn{1}{c}{ PDDE} \\[0.1cm]

\hline 
Planck\_high-$\ell$\_TTTEEE\_lite      &$585.005$&$582.857$
\\[0.1cm]
Planck\_low-$\ell$\_EE      &$396.53$&$396.29$    \\[0.1cm]
Planck\_low-$\ell$\_TT      &$ 22.75$&$23.45$    \\[0.1cm]
Planck\_lensing   &$ 8.84$&$8.77$     \\[0.1cm]
bao\_boss\_dr12          &$3.38$&$4.025$    \\[0.1cm]
bao\_smallz\_2014        &$1.99$&$2.64$  \\[0.1cm]
Pantheon        &$1025.65$&$1026.7$   \\[0.1cm]
$M_B$ prior      &$16.29$&$12.7$   \\[0.1cm]
\hline
\hline

$\chi^2_{tot}$       &$2060.47$&$2057.46$ 
\\[0.1cm]
$\bigtriangleup \chi^2_{tot}$    &$0$& $-3.01$
\\[0.1cm]
$AIC$       &$2080.47$& $2079.46$ 
\\[0.1cm]
$\bigtriangleup AIC$       &$0$& $-1.01$ 
\\[0.1cm]

\hline        
\end{tabular}
\end{table*}
To combine the SH0ES results with the other cosmological data, we take into account the SN Ia peak absolute magnitude $M_B$ rather than the $H_0$  parameter. For this, we introduce a prior on $M_B$ from the SN measurements, $M_B=-19.244\pm 0.037$. In Fig. \ref{f3},  we show the 2D and 1D posterior distributions at $68.3\%$ and $95.4\%$ C. L. for all cosmological parameters of the $\Lambda$CDM and PDDE models. The mean values, the error at $68\%$ C.L.  and $\chi^2$ per experiment are given in Table \ref{5} and Table \ref{6}, respectively. When adding the $M_B$ prior to Planck18+BAO+Pantheon, the $\Omega_{pdde}$ parameter reaches the value $0.1087_{-0.061}^{+0.052}$ and the Hubble rate increases to $H_0=69.76_{-0.82}^{+0.75}$. This increase can be observed also for the absolute magnitude where $M_B=-19.37_{-0.018}^{+0.017}$, compared to the same datasets without $M_B$ prior, as can be noticed from the strong positive
correlation in the \{$\Omega_{pdde}$, $H_0$\}, \{$\Omega_{pdde}$, $M_B$\} and \{$H_0$, $M_B$\} plans (see Fig. \ref{f3}). We also notice that the $H_0$ tension is reduced to a lower value of about $\sim 2.27 \sigma$ and the $M_B$ tension reduced to  $\sim 3.06 \sigma$ for the PDDE model.  For $\Lambda$CDM, we obtain $H_0=68.58_{-0.44}^{+0.43}$ and $M_B=-19.39\pm{0.012}$   with a tension of about $3.36\sigma$  and $3.7\sigma$, respectively. Therefore, we conclude that the PDDE model is able to  make a slight attenuation of the $H_0$ and $M_B$ tensions using Planck18+BAO+Pantheon+$M_B$ datasets compared to $\Lambda$CDM. On the other hand, the prior on $M_B$ reduces the value of $S_8$ to $0.8212\pm{0.011}$ ($0.8269_{-0.012}^{+0.011}$) for $\Lambda$CDM (PDDE), respectively, compared to the same datasets without $M_B$ prior. A negative correlation can also be seen in Fig.\ref{f3} between $S_8$ and $M_B$. According to KV450+DES-Y1,  the $S_8$ tension is at $2.16\sigma$ and $2.37\sigma$ for $\Lambda$CDM and PDDE, respectively. We notice that the $\Lambda$CDM model, reduces the $S_8$ tension compared to PDDE when constrained by Planck18+BAO+Pantheon+$M_B$ datasets.
Table \ref{6} shows the  $\chi^2$ per experiment using Planck18+BAO+Pantheon+$M_B$ datasets. We get a negative value for $\bigtriangleup\chi_{tot}^2$ and $\bigtriangleup AIC$, i. e. $\bigtriangleup\chi_{tot}^2=-3.01$ and $\bigtriangleup AIC=-1.01$, while in the previous section, positive values were found for the same datasets without $M_B$ prior. The negative value of $\bigtriangleup\chi_{tot}^2$ is mainly related to the $M_B$ prior from SH0ES data with $\bigtriangleup\chi_{M_B}^2=-3.59$. Consequently the PDDE model provides a slightly better fit for Planck18+BAO+Pantheon+$M_B$ datasets than $\Lambda$CDM. \\

\section{Effect on the CMB power spectrum.}\label{sec4}

\begin{figure*}
\centering
\includegraphics[width=18cm,height=7.5cm]{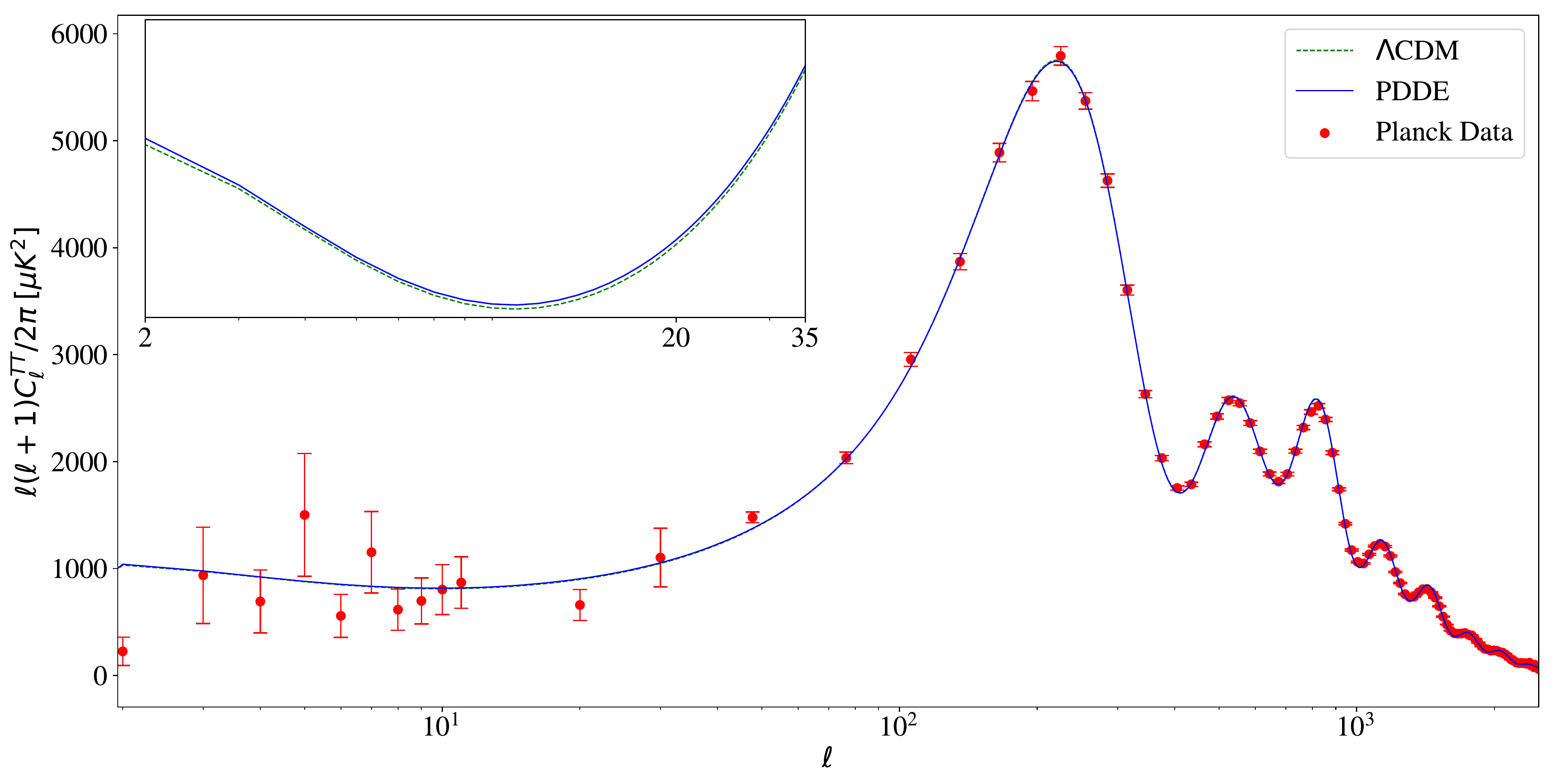} \hfill
\includegraphics[width=8.9cm,height=7.5cm]{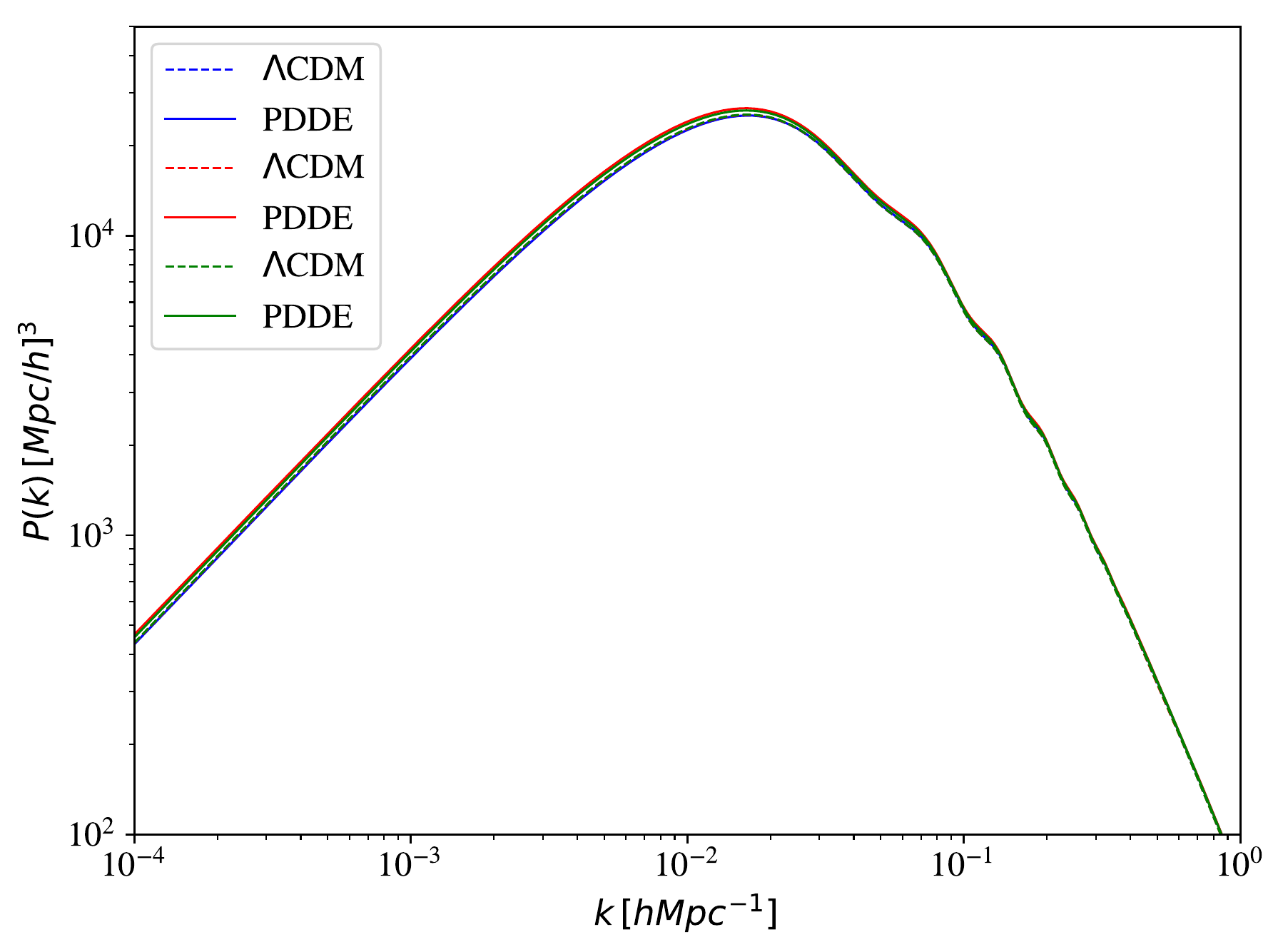}\hfill
\includegraphics[width=8.9cm,height=7.5cm]{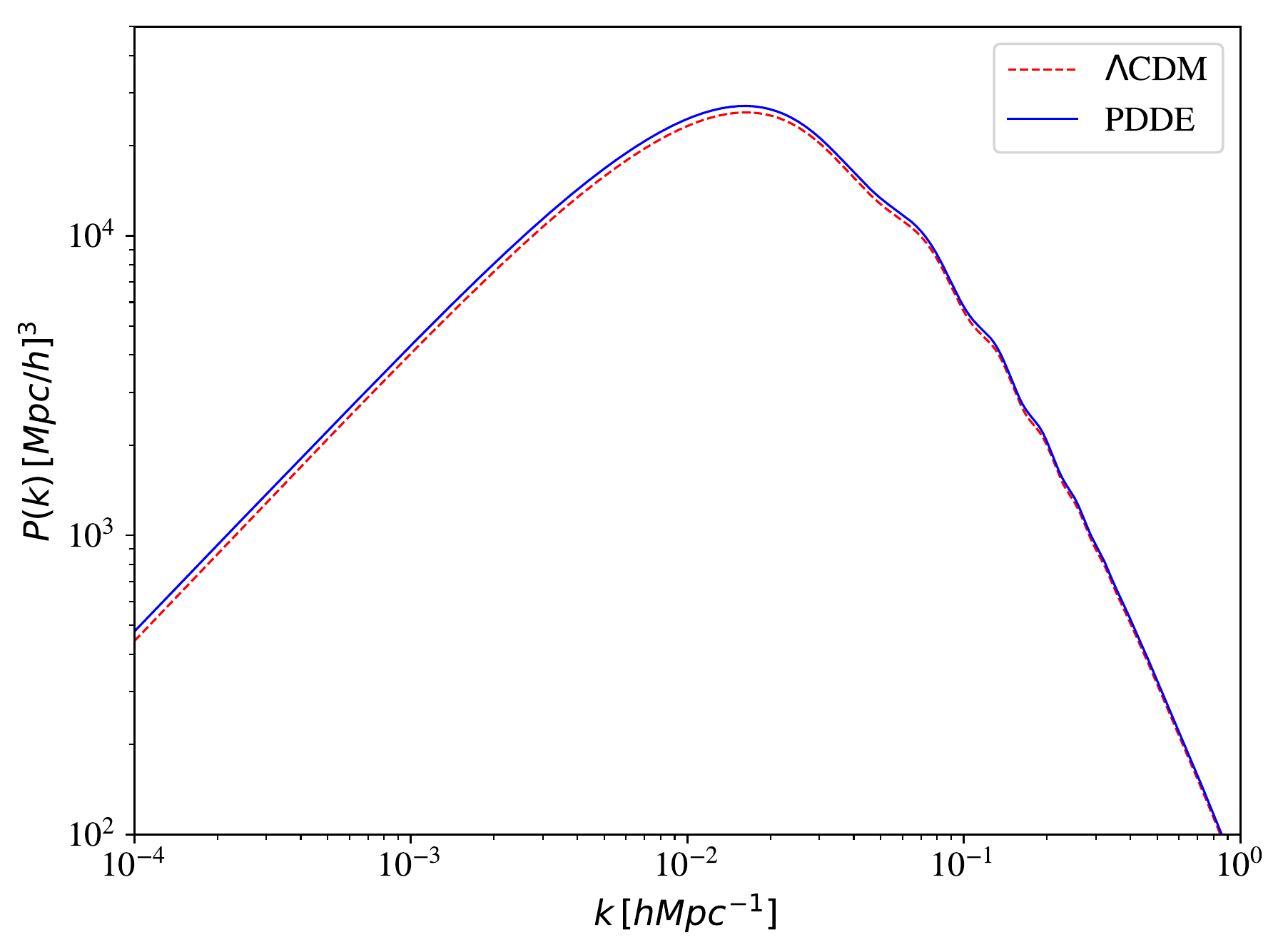}
\caption{The CMB Temperature power spectrum (top panel) for the  $\Lambda$CDM (dashed line) and PDDE (continuous line) models using the best-fit obtained by Planck18+BAO+Pantheon+$M_B$ datasets. The bottom panels correspond to the matter power spectrum using different combinations of data. The left panel is for Planck18 (blue lines), Planck18+BAO (red lines) and Planck18+BAO+Pantheon (green lines) datasets.  The right panel is for Planck18+BAO+Pantheon+$M_B$ datasets.}
\label{M}
\end{figure*}

In the top panel of Fig. \ref{M}, we show the effect of the phantom dynamical dark energy model and the $\Lambda$CDM model on the CMB temperature power spectrum  using the results obtained by Planck18+BAO+Pantheon+$M_B$ dataset. We notice that in the CMB temperature power spectrum, the effect of the PDDE model is visible at large scales $2<\ell<30$ but at higher multipoles $\ell$ this model is indistinguishable from $\Lambda$CDM. This conclusion agrees with that of  several model of this type of dark energy (see for example \cite{P6,P9}). We also show the current matter power spectrum $\mathcal{P}(z)$, for the $\Lambda$CDM and the PDDE models for different values of $\Omega_{pdde}$  using the results obtained in Tab. \ref{2} and Tab. \ref{3}. The bottom left and right panels of Fig. \ref{M} correspond to the amplitude of the matter power spectrum for different $k$-modes running approximately from  the current Hubble horizon, $k=3.33\times 10^{-4}h$ Mpc$^{-1}$ to $k\sim 1 h$ Mpc$^{-1}$. The bottom panels of Fig. \ref{M} are obtained using datasets under consideration without $M_B$ prior (left panel) and with $M_B$ prior (right panel). In the bottom-left panel of Fig. \ref{M},  we note that our phantom model is indistinguishable from $\Lambda$CDM, when  using Planck18, Planck18+BAO, Planck18+BAO+Pantheon and Planck18+BAO+Pantheon+$M_B$ datasets. Therefore, the impact of our model on the matter power spectrum is insignificant. A slightly difference observed between $\Lambda$CDM and PDDE on the amplitude of the matter power spectrum is observed in the range of smallest modes. This result is also shown in the reference \cite{LSBR2}.\\

\section{Conclusions}\label{sec3}
In this work, we have studied the effect of a phantom dynamical dark energy (PDDE) model on the cosmological parameters, particularly its capability of relieving the $H_0$ and $S_8$ tensions. The equation of state of this model depends on the redshift z and deviates from the $\Lambda$CDM model by a positive constant. This PDDE model is specified by introducing an abrupt event in the future labeled in the literature the Little Sibling of the Big Rip as it smooths the big rip singularity.
The Boltzman code CLASS has been modified to implement the parameter characterizing the PDDE model and a first Markov Chain Monte Carlo  analysis has been performed using the dataset combinations Planck18, Planck18+BAO and Planck18+BAO+Pantheon. We have found that when using Planck18+BAO datasets, a misleading reduction of the tension is noticed. In fact, finding a late-time solution of the $H_0$ tension implies an analysis of the SN measurements, i.e. Pantheon data.  Adding Pantheon
data shows a persistent $3\sigma$ tension for $H_0$ and $2.6\sigma$ for $S_8$. Although the $H_0$ tension for PDDE is reduced in comparison with $\Lambda$CDM, it is clear that a late-time model can not lead to a solution to this $H_0$ discrepancy.\\
In a second analysis, we have added a prior on  $M_B$ that was obtained by the SH0ES project, i.e. $M_B = -19.244\pm 0.037$ mag. As shown in Table \ref{5}, the PDDE model reduces the $H_0$ tension to $2.27\sigma$ and the $S_8$ tension to $2.37\sigma$ when combining Planck18+BAO+Pantheon datasets with the $M_B$ prior, i.e Planck18+BAO+Pantheon+$M_B$. Furthermore, the PDDE model provides a slightly better fit to Planck18+BAO+Pantheon+$M_B$ datasets with $\Delta\chi^2 = -3.01$ and $\Delta AIC = -1.01$ (see Table \ref{6}). \\
The distinction of the PDDE model over the standard model of cosmology is clearly observed in our work, for a wide range of data combinations. These findings agree with the fact that phantom dark energy models are supported by observations and can be an alternative of $\Lambda$CDM to solve problems related to the fine-tuning, the coincidence and the tensions under consideration if more investigations with regards to these models are done. Particularly, other phantom dark energy models such as the little rip \cite{LR} can be employed and many scenarios for the structure of the Universe such as the inclusion of massive neutrinos and the modification of the space-time curvature can be tested. We will focus on these subjects in our future works.\\
\FloatBarrier

\end{document}